# Turnover-Adjusted Information Ratio

Feng Zhang, Xi Wang and Honggao Cao[1]
Wells Fargo & Company
March 31, 2021

## Abstract

In this paper, we study the behavior of information ratio (IR) as determined by the fundamental law of active investment management. We extend the classic relationship between IR and its two determinants (i.e., information coefficient and investment "breadth") by explicitly and simultaneously taking into account the volatility of IC and the cost from portfolio turnover. Through mathematical derivations and simulations, we show that – for both mean-variance and quintile portfolios – a turnover-adjusted IR is always lower than an IR that ignores the cost from turnover; more importantly, we find that, contrary to the implication from the fundamental law but consistent with available empirical evidence, investment managers may improve their investment performance or IR by limiting/optimizing trade or portfolio turnover.

## 1. Introduction

An investment manager's performance is often evaluated by two interrelated metrics. The first is about the efficacy of the stock[2] selection model (or alpha model) that the manager uses to support her investment choices. In this, the ability of the model to distinguish between good and bad stocks, or between stocks with positive (higher) and negative (lower) excess return, is measured as information coefficient (IC). Determined cross-sectionally at a given point of time across the stocks in a targeted investment universe, IC is the correlation between the predicted excess returns and the subsequently realized excess returns of the stocks. The higher IC, the more powerful the alpha model.

The manager uses the alpha model to execute her investment strategies across time, taking into account market and non-market-related constraints, while also complementing the model-based investment decisions with qualitative judgment and discretions. The performance of the resulting investment portfolio is measured across time as the ratio between the portfolio excess return and volatility of the excess returns. This second measure is termed Information Ratio (IR). Figure 1 illustrates the mechanical relationship between IC and IR.

The relationship between IC and IR has been well studied. The best known relationship has been expressed as an element in the Fundamental Law of Active Investment Management [1]. According to this "law", IR is connected to IC through a term called "breadth" (*BR*), which is the number of independent forecasts of exceptional returns made by the manager each year. Specifically, the law states:

$$IR = IC * \sqrt{BR} \qquad (1)$$

In this classic, theoretical relationship, the performance of an investment portfolio depends on not only the power of the stock selection model used to support the investment strategies, but also the extent of portfolio "turnover", as independent forecasts typically imply investment actions or portfolio calibrations, which in turn involve moving investments into and/or out of the portfolio. In order to win in the investment management game, the portfolio manager must play often ( [2]), suggests the fundamental law. However, empirical studies have found that portfolio turnover is either irrelevant to portfolio performance ( [3]) or even negatively related to it ( [4] and [5]): the more one plays, the worse the resulting performance.

---

[1] Opinions expressed in this paper do not necessarily reflect those of Wells Fargo & CO. Correspondence should be addressed to Honggao Cao at honggao.cao@wellsfargo.com.
[2] In this paper, we use a generic term "stocks" to indicate securities covered in an alpha model, and the terms "stock" and "security" are used interchangeably.

**Figure 1. Relationship between IC and IR**

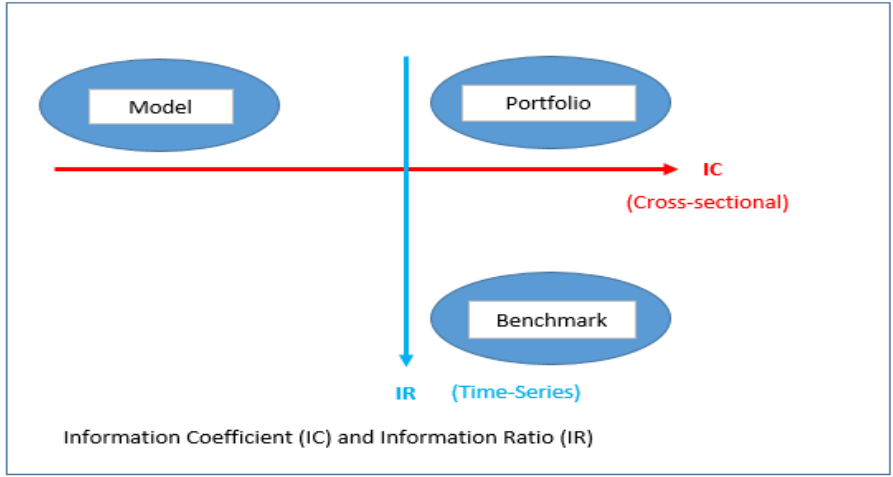

The disconnect between the theory and the reality is at least partly caused by two limitations in the IC-IR relationship as expressed in Equation (1). The first is an assumption of a constant IC over time. But the information coefficient has shown to be volatile ( [6] and [7]); the efficacy of a stock selection model may change over time. Using a constant IC to measure portfolio performance fails to account for the potential negative impact of IC volatility. The second limitation is that the IC-IR relationship does not consider transaction cost associated with portfolio turnover. Because transaction cost eats into investment returns, portfolio turnover can negatively affect portfolio performance ( [8]).

In an attempt to address the first limitation, Qian and Hua ( [6]) introduced a modified version of the fundamental law by directly recognizing the randomness of IC, which they referred to as "strategy risk". In their alternative formulation, IR is found to be the ratio of the expected value of IC relative to the IC volatility:

$$IR = \frac{\mu_{IC}}{V_{IC}} \qquad (2)$$

Here, $\mu_{IC}$ is the expected value of IC, and $V_{IC}$ is the standard deviation of IC.

The Qian-Hua formulation was subsequently expanded by Ding and Martin ( [7]), who considered the forecast error of conditional IC and provided a more general form of fundamental law of active investment management:

$$IR = \frac{\mu_{IC}}{\sqrt{V_{IC}^2 + (1-\mu_{IC}^2-V_{IC}^2)/N}} \qquad (3)$$

Here, $N$ is the number of stocks in the investment universe. Clearly, Equation (3) can be seen as a general form of Equation (2), which sets the upper bound of IR.

For the second limitation, the impact of portfolio turnover (or transaction cost) was studied by linking portfolio turnover with the deterioration of the information (or the signal) that is conveyed in the alpha model. In particular, building on the results by Qian, Sorensen and Hua ( [9] ), Ding, Martin and Yang ( [10]) showed that portfolio turnover for a mean variance portfolio is driven a so-called "alpha signal decay", which was defined as one minus the autocorrelation of the alpha signal. The turnover rate (TR) for such a portfolio can be found as:

$$TR = E_{cs}\left(\frac{1}{\sigma_i}\right) * TE * \frac{1}{\sqrt{\pi}\sqrt{V_{IC}^2 + (1-\mu_{IC}^2 - V_{IC}^2)/N}} * \sqrt{decay} \tag{4}$$

Here, $E_{cs}$ is a functional denoting cross-sectional arithmetic average; $\sigma_i$ is the security-specific volatility; $TE$ is the target portfolio tracking error; and $decay$ is the alpha signal decay as defined previously. $E_{cs}\left(\frac{1}{\sigma_i}\right)$ is the cross-sectional expectation of inverse specific volatility.

Because portfolio turnover is closely tied to "the breadth" term in the classic IC-IR relationship, the explicit link between portfolio turnover and alpha signal decay as seen in Equation (4) introduces the signal decay as another important factor into the IC-IR relationship.

The above two lines of research around the limitations in the classic IC-IR relationship, however, are not directly connected. No attempt has been made to combine the separate research efforts, or to develop an integrated understanding of information ratio that explicitly and *simultaneously* takes into account IC volatility and portfolio turnover (or signal decay).

In addition, all the aforementioned studies are conducted for mean-variance portfolios, which are typically constructed by maximizing the expected portfolio return while subject to constraints on the portfolio risk. No attempt has been made to cover another type of commonly used portfolios ( [11]): the quintile portfolios. In contrast to a mean-variance portfolio, a quintile portfolio is constructed by equally longs the top 20% (and shorts the bottom 20%) of stocks based on a stock ranking model. Quintile portfolio returns are often used as a metric in evaluating stock ranking model performance ( [12] [13]).

In this paper, we attempt to fill the above research gap. Specifically, we first extend the general form of the fundamental law of active investment management as described in Equation (3), and derive a turnover-adjusted information ratio for the mean-variance portfolios that explicitly accounts for signal decay. We then expand our research into quintile portfolios and formulate the relationship between IC and IR for such portfolios. Our alternative version of IR results may be leveraged to help the performance evaluation of both the mean-variance and quintile portfolio-based investment strategies. Finally, we also consider the impact of portfolio turnover as part of alpha signal construction, adding lagged signals into the construction process.

Through mathematical derivations and simulations, we show that portfolio managers can limit trading/turnover to improve IR, contrary to the implication embedded in the classic IC-IR relationship, that portfolio managers must play often to win in the investment management game.

## 2. Turnover-Adjusted Information Ratio for Mean-Variance Portfolios

### *Theoretical value*

Consider a mean-variance investment strategy with a target tracking error, $TE$. For this strategy, the risk-adjusted excess return may be expressed as a function of IC and alpha signal ( [10]):

$$y_{i,t} = \frac{r_{i,t}}{\sigma_i} = IC_t * x_{i,t} + \varepsilon_{i,t} \tag{5}$$

Here, $y_{i,t}$ is the risk adjusted return, which has standard deviation of 1 and mean of zero; $r_{i,t}$ is the excess return of a security $i$ at time $t$; $\sigma_i$ is the security-specific volatility; $x_{i,t}$ is the alpha signal for $i$ at $t$ (predicted for time $t$), which is standardized to have a mean of zero and a standard deviation of one; and $\varepsilon_{i,t}$ is the random error, orthogonal to $x_{i,t}$.

The risk-adjusted return is assumed to follow a standard normal distribution; this assumption is appropriate as short term security returns in an investment universe are often found to be approximately normally distributed ( [14] ). The alpha signal $x_{i,t}$ also follows a standard normal distribution, because practitioners typically normalize the returns when deriving the signals.

Information coefficient $IC_t$ is the cross-sectional correlation between the predictor $x_{i,t}$ and the risk-adjusted excess return, $\frac{r_{i,t}}{\sigma_i}$. The time series of $IC_t$ is assumed to be a stationary process ( [15]) and have a mean of $\mu_{IC}$ and a standard deviation of $V_{IC}$ as in Equation (2).

For this mean-variance portfolio, the portfolio active weight can be shown as ( [10] ):

$$w_{i,t} = \frac{TE}{\sqrt{V_{IC}^2 + (1 - \mu_{IC}^2 - V_{IC}^2)/N}} * \frac{x_{i,t}}{N * \sigma_i} \tag{6}$$

Portfolio turnover depends on $|w_{i,t-1} - w_{i,t-2}|$, which is proportional to $|x_{i,t-1} - x_{i,t-2}|$, or the absolute value of the change in portfolio active weight before and after a portfolio rebalancing. Because the alpha signal $x_{i,t}$ is a standard normal variable, the expected value of $|x_{i,t-1} - x_{i,t-2}|$ is a function of alpha signal decay, which is 1 minus the autocorrelation of the signal. Formally, the portfolio turnover rate can be shown as in Equation (4).

Given Equation (6) and the assumption that the alpha signal follows a standard normal distribution, the expected excess return of the mean-variance portfolio can then be written as:

$$E(R) = \frac{\mu_{IC}}{\sqrt{V_{IC}^2 + (1 - \mu_{IC}^2 - V_{IC}^2)/N}} * TE \tag{7}$$

In addition, the active risk of the portfolio is found to be the target tracking error, $TE$ ( [10]) .

Assuming that transaction cost ($Tcost$), the cross-sectional expectation of the inverse specific volatility ($E_{cs}\left(\frac{1}{\sigma_i}\right)$) and alpha signal decay are all constant over time, not contributing to the portfolio expected active risk, the turnover-adjusted IR is then:

$$IR' = \frac{\mu_{IC} - 2 * Tcost * E_{cs}\left(\frac{1}{\sigma_i}\right) * \sqrt{\frac{decay}{\pi}}}{\sqrt{V_{IC}^2 + (1 - \mu_{IC}^2 - V_{IC}^2)/N}} \tag{8}$$

A constant transaction cost is a reasonable and practical assumption. The constant alpha signal decay assumption will be further discussed in the paper later. For the assumption on cross-sectional specific volatility, evidence has shown that cross-sectional average specific volatility is roughly stable through time, fluctuations observed has been shown to be an episodic phenomenon ( [16]).

Notice that the turnover-adjusted IR (or $IR'$) from Equation (8) is lower than the IR from Equation (3) due to the drag of transaction cost. When transaction cost is high enough, it is possible that the turnover-adjusted IR ($IR'$) becomes negative.

### *Simulation*

The above theoretical result for turnover-adjusted IR may be verified. A simulation exercise is conducted to demonstrate this.

The following procedures are used for the simulation:

1. Simulate a normally distributed data series for information coefficient ($IC_t$). This data series covers 600 time periods (for $t$ =1, 2, …, 600), with both the mean and the standard deviation of the $IC_t$ set at 0.05.
2. Simulate a log-normally distributed data set for security-specific volatilities ($\sigma_i$). This data set covers a cross-section of 5000 securities (for $i$ = 1, 2 … 5000), and forms the target investment universe. These same security-specific volatilities will be used across time $t$.
   2a. Based on the security-specific volatility data, construct a transformed data series for inverse volatility $\frac{1}{\sigma_i}$.
3. Simulate alpha signal ($x_{i,t}$) for each of the securities in the target investment universe. This is done for each of the 600 time periods via the following scheme.
   3a. For $t$=0, simulate a standard normal data set of the alpha signal for each of the 5000 securities.
   3b. For $t$=1 to 600, simulate the alpha signal data set with period-to-period autocorrelations $(1 - decay)$ ranging from 0.6 to 0.95 with a 0.05 interval. The signal at $t + 1$ is generated based on the Equation $x_{i,t+1} = (1 - \text{decay}) * x_{i,t} + \zeta_{i,t}$. Here, $\zeta_{i,t}$ is drawn from a random normal distribution with mean of zero and variance of $1 - (1 - decay)^2$.
4. For each time period from $t$=1 to 600 and each security from $i$ =1 to 5000, generate the realized returns $r_{i,t}$ using the simulated alpha signal $x_{i,t}$, $IC_t$ and specific volatility $\sigma_i$, based on Equation (5).
5. For each time period and each security, compute the security weight $w_{i,t}$ in a mean-variance portfolio based on Equation (6), assuming that the target tracking error ($TE$) is 5%.
6. For each time period from $t$=1 to 600, calculate the simulated portfolio return as $\sum_{i=1}^{5000} w_{i,t} * r_{i,t}$, the portfolio turnover rate as $0.5 * \sum_{i=1}^{5000} abs(w_{i,t} - w_{i,t-1})$, and the portfolio return after transaction fee as $\sum_{i=1}^{5000} w_{i,t} * r_{i,t} - Tcost * \sum_{i=1}^{5000} abs(w_{i,t} - w_{i,t-1})$, assuming a transaction cost of 1%.
7. Calculate the simulated $IR$, $IR'$ and mean $TR$ based on the portfolio returns, turnover rates and other relevant quantities across the 600 time periods.
8. Repeat steps 1 to 7 for 1000 times, calculate the mean values of $IR$, $IR'$ and $TR$ and compare them with the theoretical values in Equations (3), (8) and (4).

The results from this simulation are shown in Table 1 below. They clearly indicate that for a mean-variance portfolio, all the three theoretical measures of interest, including our turnover-adjusted information ratio ($IR'$), can be properly recovered through simulation. The results also confirm that turnover-adjusted IR (or $IR'$) is always lower than the unadjusted IR, due to transaction cost.

Table 1. Theoretical and simulated portfolio Information Ratio

| | | Theoretical values | | | Simulated values | | |
|---|---|---|---|---|---|---|---|
| Autocorrelation | Decay | IR | $IR'$ | TR | IR | $IR'$ | TR |
| 0.6 | 0.4 | 0.962 | 0.666 | 0.741 | 0.964 | 0.668 | 0.741 |
| 0.65 | 0.35 | 0.962 | 0.685 | 0.693 | 0.964 | 0.687 | 0.693 |
| 0.7 | 0.3 | 0.962 | 0.706 | 0.641 | 0.964 | 0.707 | 0.641 |
| 0.75 | 0.25 | 0.962 | 0.728 | 0.586 | 0.963 | 0.729 | 0.586 |
| 0.8 | 0.2 | 0.962 | 0.753 | 0.524 | 0.964 | 0.754 | 0.524 |
| 0.85 | 0.15 | 0.962 | 0.781 | 0.454 | 0.962 | 0.781 | 0.454 |
| 0.9 | 0.1 | 0.962 | 0.814 | 0.370 | 0.962 | 0.814 | 0.370 |
| 0.95 | 0.05 | 0.962 | 0.858 | 0.262 | 0.962 | 0.857 | 0.262 |

## 3. Turnover-Adjusted Information Ratio for Quintile Portfolios

We now extend our discussion to the quintile portfolios. Here we focus on a long-short quintile portfolio. In this investment strategy, the portfolio is constructed by longing the top 20% and shorting the bottom 20% of securities from a target investment universe, based on the alpha signal $x_{i,t}$. Securities whose alpha signals are in the middle 60% of the distribution is not shorted or longed. The security weight in the portfolio is 5/N for each longed security and -5/N for each shorted security, where N is the number of the securities in the target investment universe. Long only quintile portfolios would follow a similar logic in portfolio construction as well as security weighting.

*Theoretical value*

As shown in Appendix A, the IR for a quintile portfolio with the turnover effect ignored (or the "turnover-neutral" IR) can be written as:

$$IR = \frac{2.8*\mu_{IC}*E_{cs}(\sigma_i)}{\sqrt{7.84*\sigma_{IC}^2*E_{cs}^2(\sigma_i)+E_{cs}(\sigma_i^2)*\frac{10-7.8*V_{IC}^2-7.8*\mu_{IC}^2}{N}}} \tag{9}$$

Equation (9) is very similar to Equation (3) in structure. In general, the turnover-neutral IR for a quintile portfolio is always lower than its counterpart for a mean variance portfolio; see appendix A for the comparison between Equations (3) and (9). It can also be shown that when N or the number of securities in the investment universe is large enough, both Equation (9) and Equation (3) will be effectively reduced to Equation (2).

The turnover rate (*TR*) for a quintile portfolio can be determined via the cumulative density function of a joint normal distribution by considering the fact that both the alpha signals at time $t-1$ and at time $t$ are normally distributed. For each security in the investment universe, there are 3 scenarios where one will need to buy the security.

1. The security is previously shorted but needs to be longed. In this case, the previous-period alpha signal $x_{i,t-1}$ is in the bottom 20% while the current-period alpha signal $x_{i,t}$ is in the top 20%. The probability of this situation is denoted as P1, which can be calculated using cumulative density function of the joint normal distribution of the alpha signals. In order to achieve a security weight of 5/N in the portfolio, one needs to increase the weight of the security by 2 *(5/N).

2. The security is previously not shorted or longed but needs to be longed. In this case, the previous-period alpha signal $x_{i,t-1}$ is in the middle 60% while the current-period alpha signal $x_{i,t}$ is at the top 20%. The probability of this situation is denoted as P2. In order to achieve a security weight of 5/N in the portfolio, one needs to increase the weight of the security by (5/N).

3. The security is previously shorted but needs to be dropped in the portfolio. In this case, the previous-period alpha signal $x_{i,t-1}$ is in the bottom 20% while the current-period alpha signal $x_{i,t}$ is in the middle 60%. The probability of this situation is denoted as P3. In order to achieve a security weight of 0 in the portfolio, one needs to increase the weight of the security by (5/N).

Let the autocorrelation between the alpha signal $x_{i,t-1}$ and $x_{i,t}$ be ρ. The joint cumulative density function (CDF) for the alpha signals for the two periods is then given by:

$$P(lo^{t-1} < x_{t-1} < up^{t-1}, lo^t < x_t < up^t) = \int_{lo^{t-1}}^{up^{t-1}} \int_{lo^t}^{up^t} f(x_{t-1}, x_t, \rho) \, dx_{t-1} dx \tag{10}$$

Since each long (short) security has a weight of 5/N (-5/N), the turnover rate for the portfolio is

$$TR = \frac{5}{N} * N * (2*P1 + P2 + P3) = 5*(2*P1 + P2 + P3) \tag{11}$$

The turnover rate as a function of decay $(1 − ρ)$ is plotted in Figure 2.

**Figure 2. Turnover rate as a function of decay for quintile portfolio**

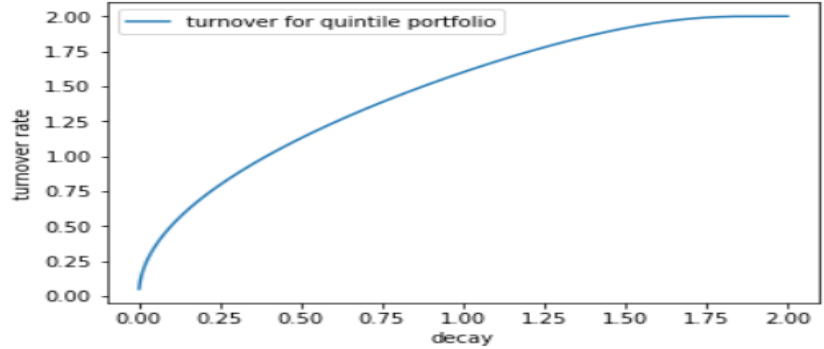

The turnover-adjusted IR for the quintile portfolio is:

$$IR' = \frac{2.8*\mu_{IC}*E_{cs}(\sigma_i) - 2*Tcost*TR}{\sqrt{7.84*V_{IC}^2*E_{cs}^2(\sigma_i) + E_{cs}(\sigma_i^2)*\frac{10-7.8*V_{IC}^2-7.8*\mu_{IC}^2}{N}}} \quad (12)$$

*Simulation*

To demonstrate the relevance of the turnover-adjusted IR for quintile portfolios, we conduct a simulation exercise, mimicking the simulation procedures as described in section 2. The only difference in this exercise is that the security weight is now the weight for a quintile portfolio (e.g., 5/N, -5/N, or 0).

The simulation results are shown in Table 2 below. Again, they clearly indicate that for a quintile portfolio, all the three theoretical measures of interest, including our turnover-adjusted information ratio, can be properly recovered through simulation.

**Table 2. Theoretical and simulated portfolio Information Ratio for quintile portfolio**

|  |  | Theoretical values | | | Simulated values | | |
| --- | --- | --- | --- | --- | --- | --- | --- |
| Autocorrelation | Decay | IR | $IR'$ | TR | IR | $IR'$ | TR |
| 0.6 | 0.4 | 0.948 | 0.680 | 1.008 | 0.950 | 0.681 | 1.008 |
| 0.65 | 0.35 | 0.948 | 0.698 | 0.942 | 0.952 | 0.701 | 0.942 |
| 0.7 | 0.3 | 0.948 | 0.717 | 0.871 | 0.948 | 0.716 | 0.871 |
| 0.75 | 0.25 | 0.948 | 0.737 | 0.794 | 0.948 | 0.737 | 0.794 |
| 0.8 | 0.2 | 0.948 | 0.759 | 0.710 | 0.949 | 0.760 | 0.710 |
| 0.85 | 0.15 | 0.948 | 0.785 | 0.614 | 0.949 | 0.786 | 0.614 |
| 0.9 | 0.1 | 0.948 | 0.815 | 0.501 | 0.951 | 0.817 | 0.501 |
| 0.95 | 0.05 | 0.948 | 0.854 | 0.354 | 0.948 | 0.854 | 0.354 |

## 4. Turnover-Adjusted Information Ratios: Mean-Variance versus Quintile Portfolios

To compare the turnover-adjusted IR between mean-variance portfolios and quintile portfolios, we calculate the turnover-adjusted IRs at different decay factor levels for both types of portfolios. In this comparison, we assume $Tcost$=1%, $\mu_{IC}$=0.05, $V_{IC}$=0.05, and $N$=5000. In addition, we assumed that $\log(\sigma_i)$ has a normal distribution with mean -0.722 and standard deviation of 0.306. Figure 3 displays the turnover-adjusted IR as a function of decay for the two types of portfolios. It shows that the mean-variance portfolio has a slightly larger turnover-adjusted

IR when decay is small (<0.09 in this case); and the quintile portfolio has a higher turnover-adjusted IR when decay is large.

**Figure 3. Turnover-adjusted IR as a function of decay**

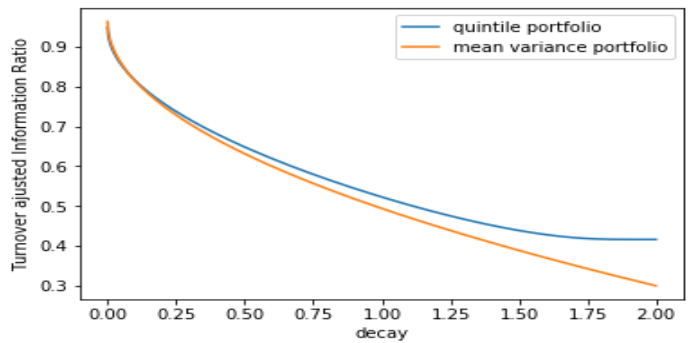

## 5. Signal decay in Turnover-adjusted Information Ratio

As discussed previously, signal decay is defined as one minus the signal autocorrelation. Signals from different alpha models can have different decay rates. Momentum factors can decay very quickly. Portfolios based on such factors tend to have high turnover. Valuation factors generally decay slowly; valuation-based portfolios tend to have low turnover. Table 3 below collects the signal decays for some factors/models based on publicly available data. It shows that most factors/models decay slowly (decay<=0.1) with the exception of momentum factors.

**Table 3. Alpha signal decay for Selected Industry Studies**

| Factor/Model | Average IC | Signal decay | Investment Universe | Study period | Model frequency |
|---|---|---|---|---|---|
| 9-month momentum | 0.073 | 0.32 | Russell 3000 | 1987-2004 | Quarterly |
| E/P | 0.043 | 0.06 | Russell 3000 | 1987-2004 | Quarterly |
| B/P | 0.029 | 0.06 | Russell 3000 | 1979-2018 | Monthly |
| C/P | 0.032 | 0.07 | Russell 3000 | 1979-2018 | Monthly |
| D/P | 0.029 | 0.03 | Russell 3000 | 1979-2018 | Monthly |
| E/P | 0.031 | 0.07 | Russell 3000 | 1979-2018 | Monthly |
| FE/P | 0.034 | 0.06 | Russell 3000 | 1979-2018 | Monthly |
| S/P | 0.020 | 0.03 | Russell 3000 | 1979-2018 | Monthly |
| MOM | 0.025 | 0.13 | Russell 3000 | 1979-2018 | Monthly |
| SHORT | 0.036 | 0.07 | Russell 3000 | 1979-2018 | Monthly |
| Growth Benchmark Model | 0.024 | 0.09 | Russell 3000 Growth | 2018 | Monthly |
| Value Benchmark Model | 0.017 | 0.08 | Russell 3000 Value | 2018 | Monthly |
| Quality Model | 0.040 | 0.10 | Russell 3000 | 2018 | Monthly |
| Price Momentum Model | 0.009 | 0.39 | Russell 3000 | 2018 | Monthly |
| Growth Benchmark Model | 0.011 | 0.08 | Russell 3000 Growth | 2016 | Monthly |
| Value Benchmark Model | 0.017 | 0.08 | Russell 3000 Value | 2016 | Monthly |
| Quality Model | 0.005 | 0.10 | Russell 3000 | 2016 | Monthly |
| Price Momentum Model | -0.001 | 0.38 | Russell 3000 | 2016 | Monthly |

Note: Data from [9], [10] [17] [18] Signal decays from [17] [18] are based on rank correlation; B/P = book to price, C/P = cash flow to price, D/P = dividend yield to price, E/P =earnings to price, FE/P = forward earnings to price, S/P = sales to price, MOM = cumulative 11-month return from t-12 to t-2, SHORT = short as a percent of total shares float

In our derivation of the turnover-adjusted information ratios (i.e., in Equations (8) and (12)), we assumed that signal decay is constant overtime. The signal decay of S&P stock selection models ( [17] [18]) from two different years are very close (with the differences less than 0.01 between 2016 and 2018), suggesting that our assumption of constant signal decay is generally valid.

**6. Integrated signals**

Combining lagged signals with the current signal to form an integrated alpha signal has been used as a method to improve the IR for investment strategies that are conscious of turnover. This is possible because an integrated signal, while less predictive for the current return, may reduce signal decay ( [9]), which in turn may reduce transaction cost. In fact, researchers have found that constructing a portfolio with the consideration of transaction cost is equivalent to constructing the portfolio using a weighted average of current and past alpha signals ( [19]).

*Integrated signals through one lag*

To show how the integrated alpha signals may affect our turnover-adjusted IR, we first consider the case where the integrated signal for a security at time *t* is just a weighted average of the current signal and the lag one signal. That is,

$$A_t = w_1 * x_t + w_2 * x_{t-1} \tag{13}$$

Here, $A_t$ is the integrated alpha signal, $x_t$ is the current alpha signal at time *t*, and $w_1 + w_2 = 1$. Let the information coefficient of the current signal $x_t$ be $IC_{x,t}$. The mean and standard deviation of $IC_{x,t}$ are denoted by $\mu_{IC}$ and $V_{IC}$, respectively. The current signal and the lagged signal $x_{t-1}$ are correlated, and the correlation coefficient is $\rho$. As a result, the IC of the lagged signal is $\rho * IC_{x,t}$. Finally, the IC for lagged signal is assumed to decay at the same rate as the current signal.

With this setup, the integrated signal $A_t$ has an IC as a function of $w_1, w_2\ IC_{x,t}$ and $\rho$, as follows:

$$IC_{A,t} = \frac{Cov(y_t, A_t)}{\sqrt{Var(A_t)}} = \frac{w_1 + w_2 * \rho}{\sqrt{w_1^2 + 2*w1*w_2*\rho + w_2^2}} * IC_{x,t} \tag{14}$$

Here, $y_t$ is the risk-adjusted return as defined in Equation (5) earlier.

The IC of the integrated signal has a mean of $\frac{w_1 + w_2 * \rho}{\sqrt{w_1^2 + 2*w1*w_2*\rho + w_2^2}} * \mu_{IC}$ and a standard deviation of $\frac{w_1 + w_2 * \rho}{\sqrt{w_1^2 + 2*w1*w_2*\rho + w_2^2}} * V_{IC}$. Both the mean and the standard deviation are scaled by the discount factor of $\frac{w_1 + w_2 * \rho}{\sqrt{w_1^2 + 2*w1*w_2*\rho + w_2^2}}$.

The autocorrelation for the integrated alpha signal is given by

$$\rho_A = \frac{w_1^2 * \rho + w1*w_2*\rho^2 + w1*w_2 + w_2^2*\rho}{w_1^2 + 2*w1*w_2*\rho + w_2^2} \tag{15}$$

The relationship between *w*1 and the turnover-adjusted IR for both mean-variance portfolios and quintile portfolios is plotted in Figure 4. In this plot, we assume $\mu_{IC}$=0.05, $V_{IC}$=0.1, *N*=5000, and $\rho$=0.9; we also assume that log(σ$_i$) has a normal distribution with mean -0.722 and standard deviation of 0.306. For the turnover cost, we choose two alternatives: *Tcost* = 1% and 3%.

**Figure 4. Turnover-adjusted IR of integrated alpha signal as function of weight in current signal**

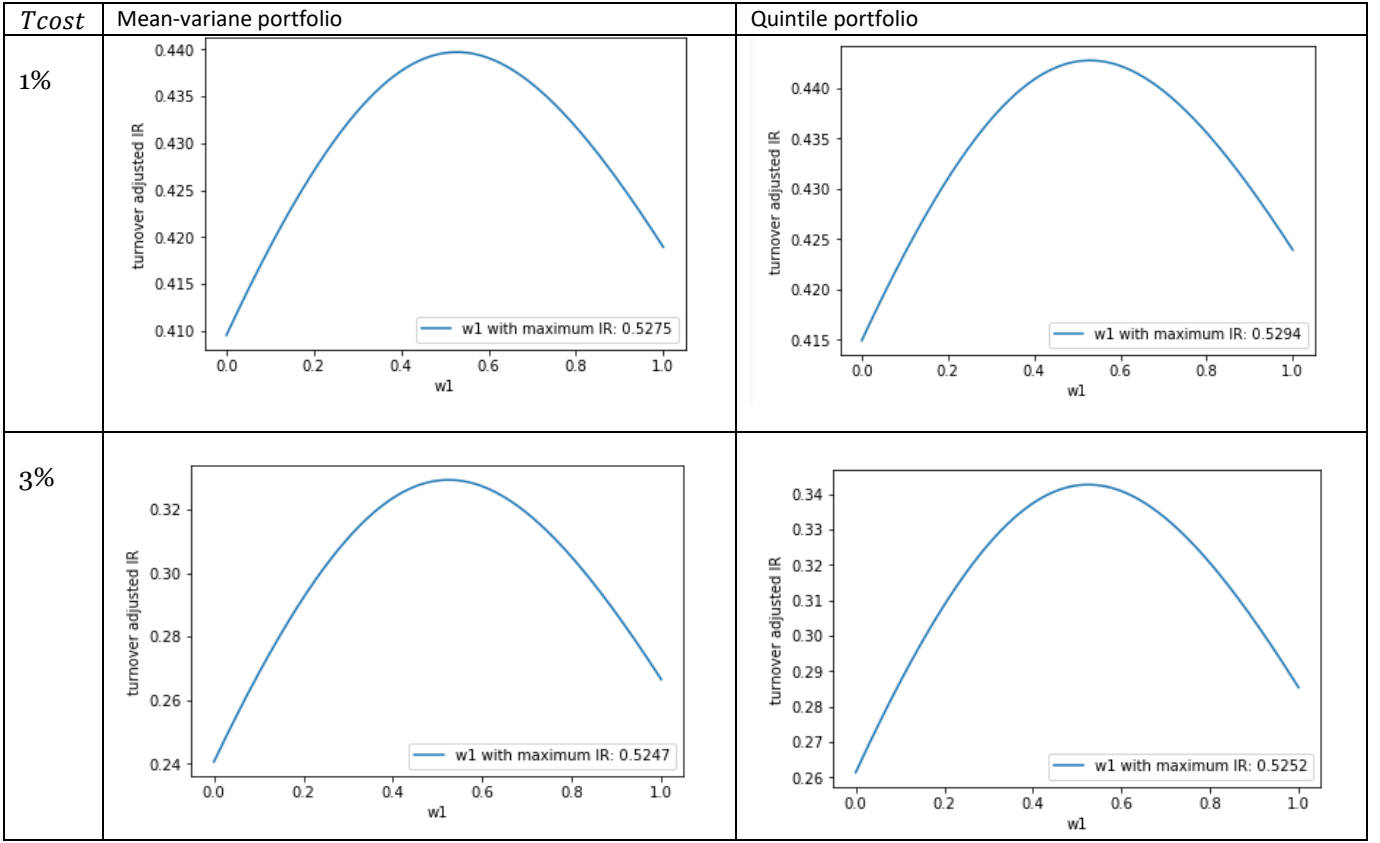

The following observations can be made from the plot. First, there exists an optimal $w1$ that can maximize the turnover-adjusted IR for both mean-variance portfolios and quintile portfolios. Second, in consistence with our expectation, higher transaction costs would require more weighting on the lagged signal in order to achieve the optimal IR. Finally, the impact of transaction cost on optimal weight is relatively small.

*Integrated signal through multiple lags*

We next extend the discussions above to consider an integrated alpha signal that is formed through multiple lags. We use an exponentially weighted moving average approach ( [19]) to combine the current signal and the lags. That is,

$$A_t = \sum_{j=0}(1 - \lambda) * \lambda^j * x_{t-j} \qquad (16)$$

Here $\lambda$ is a weighting factor with values between 0 and 1. The weights on concurrent signal equals $1 - \lambda$. A smaller $\lambda$ would weight recent signals more heavily. We maintain the assumption that the IC for lagged signal decays at the same rate as the current signal, and further assume that the alpha signals follow an AR(1) process.[3]

With the above structure, the autocorrelation for the integrated alpha signal is ( [19]):

$$\rho_A = \frac{\lambda + \rho}{1 + \lambda * \rho} \qquad (17)$$

---

[3] Actual alpha signal IC and autocorrelation might decay faster than AR(1). This does not impact our main conclusion that integrating lagged signal(s) could improve turnover-adjusted IR, although the optimal weighting on current signal vs lagged signals might be smaller than implied from maximizing Equation (20).

The variance of the integrated signal is ([20]):

$$Var(A_t) = \frac{(1-\lambda)(1+\lambda*\rho)}{(1+\lambda)(1-\lambda*\rho)} \quad (18)$$

As a result, the IC for the integrated signal $A_t$ can be determined as in Equation (19):

$$IC_{A,t} = \frac{Cov(y_t, A_t)}{\sqrt{Var(A_t)}} = \frac{1-\lambda}{1-\lambda*\rho} * \frac{1}{\sqrt{Var(A_t)}} * IC_{x,t} = \sqrt{\frac{1-\lambda^2}{1-\lambda^2*\rho^2}} * IC_{x,t} \quad (19)$$

For the mean-variance portfolios, the turnover-adjusted IR that is derived from the integrated signal is:

$$IR_{A,t}' = \frac{\mu_{IC} - 2*Tcost*E_{CS}\left(\frac{1}{\sigma_i}\right)*\sqrt{\frac{1-\rho}{\pi}}*\sqrt{\frac{1-\lambda*\rho}{1+\lambda}}}{\sqrt{V_{IC}^2 - \frac{\mu_{IC}^2 + V_{IC}^2}{N} + \frac{1-\lambda^2*\rho^2}{N(1-\lambda^2)}}} \quad (20)$$

The turnover-adjusted IR for the quintile portfolios can be calculated as a function of the integrated signal similarly.

According to Equation 20, there exists a $\lambda$ between 0 and 1 that maximizes the turnover-adjusted IR (see Appendix B for more details). In addition, if $\lambda$ get close to 1, the average IC of the integrated alpha signal will approach zero, driving the turnover-adjusted IR towards zero. If $\lambda$ get close to 1, portfolios constructed based on the integrated alpha signal would become passive portfolios that basically just follow the benchmarks. This indicates that a portfolio manager can always avoid negative IR if the performance of an alpha model is known. In an extreme situation where cost of trading is larger than the potential profit, portfolio managers may just need to sit tight and follow the benchmarks.

The relationship between $\lambda$ and turnover-adjusted IR is plotted in Figure 4. In this plot, we again assume $\mu_{IC}=0.05$, $V_{IC}=0.1$, $N=5000$, and $\rho=0.9$; we also assume that $\log(\sigma_i)$ has a normal distribution with mean -0.722 and standard deviation of 0.306. The results in Figure 5 are consistent with expectation: higher transaction costs would require more weighting on lagged signals in order to achieve optimal IR. In addition, optimal $\lambda s$ exist for both mean-variance and quintile portfolios that maximize the turnover-adjusted IRs.

**Figure 5 Turnover-adjusted IR of composite signal as function of decay factor**

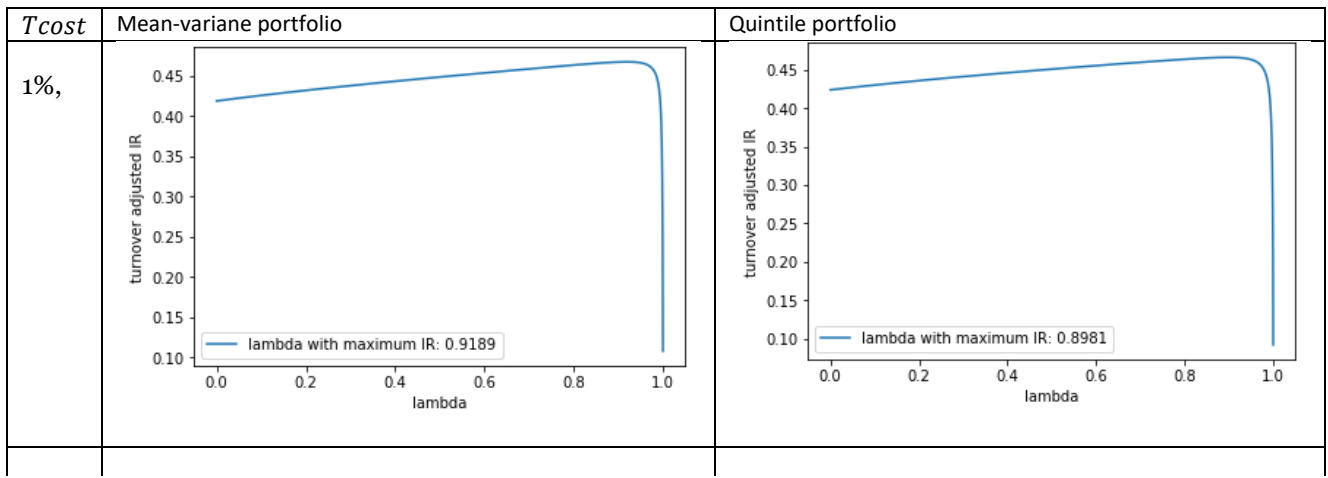

3%, 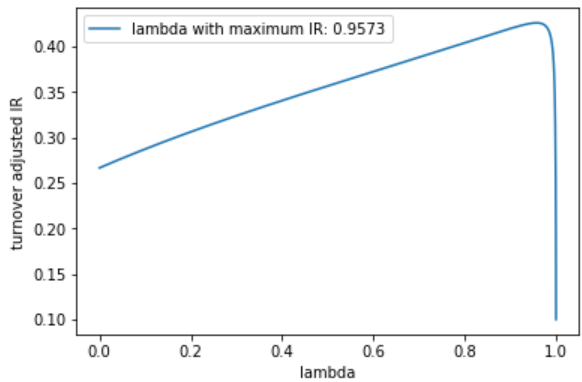 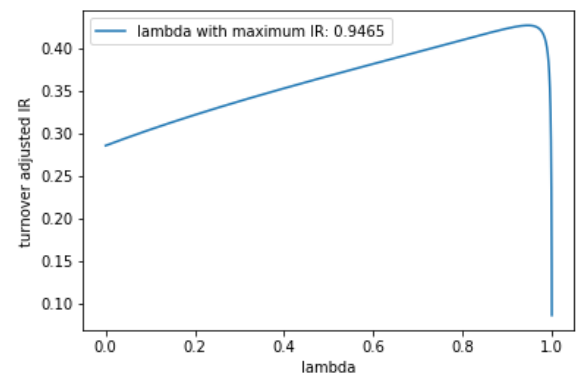

When explaining the classic, fundamental law of active management, Grinold and Kahn [2] conclude that "you (portfolio managers) must play often and play well to win at the investment management game. It takes only a modest amount of skills to win as long as that skill is deployed frequently and across a large number of stocks." However, our results above suggest that portfolio managers can limit trading/turnover to improve IR. By integrating lagged signal(s) into the current alpha signal, the turnover-adjusted IR can be improved through decreased turnover.

**6. Conclusion**

In this paper, we have extended the general form of fundamental law of active investment management to include the effect of portfolio turnover for both mean-variance portfolios and quintile portfolios. This alternative version of information ratio (IR) takes into account the mean IC, IC volatility and alpha signal decay in an integrated manner. It may be leveraged to help portfolio managers assess their alpha models and investment strategies.

We have also studied the impact of portfolio turnover as part of alpha signal construction, where the lagged signals are added in the construction process. We have shown that adding lagged alpha signal can improve the turnover-adjusted IR by limiting portfolio turnover. Our finding that portfolio managers can limit trading/turnover to improve IR is in contrary to the implication embedded in the classic IC-IR relationship, that portfolio managers must play often to win in the investment management game.

Our analysis demonstrates that if the performance of an alpha model is known and stable, negative IR can always be avoided by limiting turnover. Observed negative IRs may be a result of undetected model performance deterioration, causing excessive trading in an attempt to chase the signals that have already deteriorated or disappeared.

**Appendices**

*A. IR for Quintile portfolios*

In the main text, we provided the theoretical information coefficient formula for quintile portfolios. In this appendix, we show how the theoretical information ratio for quintile portfolios is derived.

A quintile portfolio is constructed by equally longs the top 20% (and shorts the bottom 20%) of stocks based on a stock ranking model. The return of a quintile portfolio is the equal-weighted return to the top 20% of stocks minus the equal-weighted return to the bottom 20% of stocks.

As noted in the main text, the risk adjusted excess return $\frac{r_{it}}{\sigma_i}$ can be expressed as a function of IC and alpha signal.

$$\frac{r_{i,t}}{\sigma_i} = IC_t * x_{i,t} + \varepsilon_{i,t} \tag{A1}$$

Here, $\left(\frac{r_{i,t}}{\sigma_i}\right) = 0$, $\sigma\left(\frac{r_{i,t}}{\sigma_i}\right) = 1$, $E(x_{i,t}) = 0$ and $\sigma(x_{i,t}) = 1$ by the "normalization" of the returns, and $\varepsilon_{i,t}$ is a vector of random errors, which is orthogonal to $x_t$. In this setup, the regression coefficient, $IC_t$, is the information coefficient (IC) at time $t$ of the stock selection model. $IC_t$ is assumed to follow normal distribution with mean $\mu_{IC}$ and volatility $V_{IC}$. The stock-specific volatility $\sigma_i$ is assumed to be the same across different time periods for a given security.

We can rewrite Equation A1 as Equation A2.

$$r_{i,t} = \sigma_i IC_t x_{i,t} + \sigma_i \varepsilon_{i,t} \tag{A2}$$

Based on the characteristics of truncated normal distribution, the expected return of quintile portfolio (quintile spread) is:

$$E(Quintile\ Spread) = E(IC_t) * \left(\frac{\phi(normal.inverse(0.8))}{1-0.8}\right) * E_{cs}(\sigma_i) - E(IC_t) * \left(\frac{-\phi(normal.inverse(0.2))}{0.2}\right) * E_{cs}(\sigma_{r_i})$$

$$\tag{A3}$$

$$E(Quintile\ Spread) \approx 2.8 * \mu_{IC} * E_{cs}(\sigma_i) \tag{A4}$$

Here, $\phi$ is the probability density function of the standard normal distribution.

The variance of quintile spread can be written as:

$$Var(Quintile\ Spread) = Var\left(avg(r_{i,t}|x_{i,t} > 80\ percentile) - avg(r_{i,t}|x_{i,t} < 20\ percentile)\right)$$

$$\approx Var\left(IC_t * \overline{\sigma_i x_{i,t}}_{x_{i,t} \geq Q_{0.8}} + \overline{\sigma_i \varepsilon_{i,t}}_{x_{i,t} \geq Q_{0.8}}\right) + VAR\left(IC_t * \overline{\sigma_i x_{i,t}}_{x_{i,t} \leq Q_{0.2}} + \overline{\sigma_i \varepsilon_{i,t}}_{x_{i,t} \geq Q_{0.2}}\right)$$

$$-2 * Cov(IC_t * \overline{\sigma_i x_{i,t}}_{x_{i,t} \geq Q_{0.8}} + \overline{\sigma_i \varepsilon_{i,t}}_{x_{i,t} \geq Q_{0.8}}, IC_t * \overline{\sigma_i x_{i,t}}_{x_{i,t} \leq Q_{0.2}} + \overline{\sigma_i \varepsilon_{i,t}}_{x_{i,t} \geq Q_{0.2}})$$

$$\tag{A5}$$

Here, the $\overline{\sigma_i X}_{x \geq Q_{0.8}}$ is the cross-sectional mean of $\sigma_i * x_{i,t}$ given $x_{i,t}$ at the top 20%, and $\overline{\sigma_i \varepsilon_{i,t}}_{x_{i,t} \geq Q_{0.8}}$ is the cross-sectional mean of $\sigma_i * \varepsilon_{i,t}$ given $x_{i,t}$ at the top 20%. Since $\varepsilon_{i,t}$ is independent from $x_{i,t}$, $E(\overline{\sigma_i \varepsilon_{i,t}}_{x_{i,t} \geq Q_{0.8}}) = 0$.

$$Var(IC_t * \overline{\sigma_r X}_{x \geq Q_{0.8}} + \overline{\sigma_r \varepsilon_t^{0.8}}) = V_{IC}^2 * \sigma^2_{\overline{\sigma_i x_{i,t}}_{x_{i,t} \geq Q_{0.8}}} + V_{IC}^2 * u^2_{\overline{\sigma_i x_{i,t}}_{x_{i,t} \geq Q_{0.8}}} + \mu_{IC}^2 * \sigma^2_{\overline{\sigma_i x_{i,t}}_{x_{i,t} \geq Q_{0.8}}} + \sigma^2_{\overline{\sigma_i \varepsilon_{i,t}}_{x_{i,t} \geq Q_{0.8}}} \tag{A6}$$

$$Var(IC_t * \overline{\sigma_r X}_{x \leq Q_{0.2}} + \overline{\sigma_r \varepsilon_t^{0.2}}) = V_{IC}^2 * \sigma^2_{\overline{\sigma_i x_{i,t}}_{x_{i,t} \leq Q_{0.2}}} + V_{IC}^2 * u^2_{\overline{\sigma_i x_{i,t}}_{x_{i,t} \leq Q_{0.2}}} + \mu_{IC}^2 * \sigma^2_{\overline{\sigma_i x_{i,t}}_{x_{i,t} \leq Q_{0.2}}} + \sigma^2_{\overline{\sigma_i \varepsilon_{i,t}}_{x_{i,t} \leq Q_{0.2}}} \tag{A7}$$

$$Cov(IC_t * \overline{\sigma_i x_{i,t}}_{x_{i,t} \geq Q_{0.8}} + \overline{\sigma_i \varepsilon_{i,t}}_{x_{i,t} \geq Q_{0.8}}, IC_t * \overline{\sigma_i x_{i,t}}_{x_{i,t} \leq Q_{0.2}} + \overline{\sigma_i \varepsilon_{i,t}}_{x_{i,t} \geq Q_{0.2}})$$

$$= E(IC_t^2 * \overline{\sigma_i x_{i,t}}_{x_{i,t} \geq Q_{0.8}} * \overline{\sigma_i x_{i,t}}_{x_{i,t} \leq Q_{0.2}}) - E(IC_t * \overline{\sigma_r X}_{x \geq Q_{0.8}}) * E(IC_t * \overline{\sigma_r X}_{x \leq Q_{0.2}})$$

$$= E(\text{IC}_t^2) E\left(\overline{\sigma_i x_{i,t}}_{x_{i,t} \geq Q_{0.8}}\right) E\left(\overline{\sigma_i x_{i,t}}_{x_{i,t} \leq Q_{0.2}}\right) + \mu_{IC}^2 * u_{\overline{\sigma_i x_{i,t}}_{x_{i,t} \geq Q_{0.8}}}^2$$

$$= -(\mu_{IC}^2 + V_{IC}^2) * u_{\overline{\sigma_i x_{i,t}}_{x_{i,t} \geq Q_{0.8}}}^2 + \mu_{IC}^2 * u_{\overline{\sigma_i x_{i,t}}_{x_{i,t} \geq Q_{0.8}}}^2$$

$$= -V_{IC}^2 * \mu^2(\overline{\sigma_r X}_{x \geq Q_{0.8}})$$

$$= -V_{IC}^2 * 1.96 * E_{CS}^2(\sigma_i) \tag{A8}$$

$$\sigma_{\overline{\sigma_i x_{i,t}}_{x_{i,t} \geq Q_{0.8}}}^2 = \sigma_{\overline{\sigma_i \varepsilon_{i,t}}_{x_{i,t} \leq Q_{0.2}}}^2 = \frac{\sum \sigma_i^2}{n^2} * \sigma_{\overline{x_{i,t}}_{x_{i,t} \leq Q_{0.2}}}^2 \approx \frac{0.22 * \sum \sigma_i^2}{n^2} = \frac{0.22 * E_{cs}(\sigma_i^2)}{n} \tag{A9}$$

$$\sigma_{\overline{\sigma_i \varepsilon_{i,t}}_{x_{i,t} \geq Q_{0.8}}}^2 = \sigma_{\overline{\sigma_i \varepsilon_{i,t}}_{x_{i,t} \leq Q_{0.2}}}^2 = \frac{\sum \sigma_i^2}{n^2} * \sigma_{\varepsilon_{i,t}}^2 = \frac{(1 - \sigma_{IC}^2 - IC^2) * E_{cs}(\sigma_i^2)}{n} \tag{A10}$$

$$\mu_{\overline{\sigma_i x_{i,t}}_{x_{i,t} \geq Q_{0.8}}} = -\mu_{\overline{\sigma_i x_{i,t}}_{x_{i,t} \leq Q_{0.2}}} = \frac{\sum \sigma_i}{n} \mu_{\overline{x_{i,t}}_{x_{i,t} \geq Q_{0.8}}} = E_{cs}(\sigma_i) \left(\frac{\phi(\text{normal.inverse}(0.8))}{1 - 0.8}\right) \approx 1.4 * E_{cs}(\sigma_i) \tag{A11}$$

Apply Equations of A6 to A11 to A7 and n=5/N,

$$\text{Var (Quintile Spread)} = 7.84 * V_{IC}^2 * E_{CS}^2(\sigma_i) + E_{cs}(\sigma_i^2) * \frac{10 - 7.8 * V_{IC}^2 - 7.8 * IC^2}{N} \tag{A12}$$

The information ration for quintile portfolio is:

$$IR = \frac{2.8 * IC * E_{cs}(\sigma_i)}{\sqrt{7.84 * V_{IC}^2 * E_{CS}^2(\sigma_i) + E_{cs}(\sigma_i^2) * \frac{10 - 7.8 * V_{IC}^2 - 7.8 * IC^2}{N}}} \tag{A13}$$

Equation A13 can be written as

$$IR = \frac{IC}{\sqrt{V_{IC}^2 + \frac{E_{cs}(\sigma_i^2)}{E_{CS}^2(\sigma_i)} * \frac{10 - 7.8 * V_{IC}^2 - 7.8 * IC^2}{7.84 * N}}} \tag{A14}$$

Since $E_{cs}(\sigma_i^2) >= E_{cs}^2(\sigma_i)$, and $\frac{10 - 7.8 * V_{IC}^2 - 7.8 * IC^2}{7.84 * N} > \frac{7.84 - 7.84 * V_{IC}^2 - 7.84 * IC^2}{7.84 * N}$, Equation A14 is smaller than Equation (3), meaning information ratio for quintile portfolio is always lower than mean-variance portfolio given the same investment universe and same alpha signal.

## B: Optimal turnover-adjusted IR with respect to λ

In the main text, we state that a λ between 0 and 1 can maximize the turnover-adjusted information ratios. To prove that Equation (20) has a maximum value for some λ between 0 and 1, we formulate the first derivative of Equation (20). To facilitate the derivation, we define $k = V_{IC}^2 - \frac{\mu_{IC}^2 + V_{IC}^2}{N}$, $x = \lambda$, and $m = 2 * Tcost * E_{cs}\left(\frac{1}{\sigma_i}\right) / \sqrt{\pi}$.

The first order derivative of Equation (20) equals:

$$\frac{\frac{m(1+p)\sqrt{1-p}}{2(1+x)\sqrt{1+x}\sqrt{1-xp}} \sqrt{1 - x^2 p^2 + k(N - Nx^2)} - \frac{x(1-p^2)}{\sqrt{1 + k(N - Nx^2) - x^2 p^2}} \frac{1}{\sqrt{N}(1-x^2)^{3/2}} \left(\mu_{IC} - \frac{m\sqrt{1-p}\sqrt{1-xp}}{\sqrt{1+x}}\right)}{\sqrt{\frac{1 - x^2 p^2}{N - Nx^2} + k}^2} \tag{A15}$$

The sign of the derivative is thus determined by the numerator.

$$\frac{m(1+p)\sqrt{1-p}}{2(1+x)\sqrt{1+x}\sqrt{1-xp}}\sqrt{1-x^2p^2}+k(N-Nx^2))-\frac{x(1-p^2)}{\sqrt{1+k(N-Nx^2)-x^2p^2}}\frac{1}{\sqrt{N}(1-x^2)^{3/2}}\left(\mu_{IC}-\frac{m\sqrt{1-p}\sqrt{1-xp}}{\sqrt{1+x}}\right) \quad \text{(A16)}$$

When $x \to 0$, the derivative is positive as the second part of the Equation A2 approaches zero and the first part is positive; when $x \to 1$, the derivative is negative as the second part of A2 would explode to infinity due to the $(1-x^2)^{3/2}$ part in the denominator. According to the mean value theorem, there thus exists a maximum value of Equation (20) for λ between 0 and 1.